\documentclass[showpacs,preprintnumbers,amsmath,amssymb]{revtex4}
\usepackage{amsthm}
\usepackage{enumerate}
\usepackage{epsfig}
 \newtheorem{theorem}{Theorem}[section]

 \newtheorem{definition}[theorem]{Definition}

\newcommand*{\cF}{\mathcal{F}}

\newcommand*{\cM}{\mathcal{M}}
\newcommand*{\cN}{\mathcal{N}}

\newcommand*{\cS}{\mathcal{S}}

\newcommand*{\ket}[1]{|#1\rangle}
\newcommand*{\bra}[1]{\langle #1|}

\newcommand{\ketbra}[1]{| #1 \rangle \langle #1 |}

\newcommand{\braket}[2]{\langle #1|#2\rangle}       

\newcommand{\be}{\begin{equation}}
\newcommand{\ee}{\end{equation}}
\newcommand{\bea}{\begin{eqnarray}}
\newcommand{\eea}{\end{eqnarray}}
\newcommand{\bestar}{\begin{equation*}}
\newcommand{\eestar}{\end{equation*}}
\newcommand{\beastar}{\begin{eqnarray*}}
\newcommand{\eeastar}{\end{eqnarray*}}

\begin{document}

\title{Sequential weak measurement}

\author{Graeme \surname{Mitchison}}
\email[]{g.j.mitchison@damtp.cam.ac.uk} \affiliation{Centre for
Quantum Computation, DAMTP,
             University of Cambridge,
             Cambridge CB3 0WA, UK}

\author{Richard \surname{Jozsa}}
\email[]{r.jozsa@bristol.ac.uk} \affiliation{Department of Computer
Science, University of Bristol, Bristol, BS8 1UB, UK}

\author{Sandu \surname{Popescu}}
\email[]{s.popescu@bristol.ac.uk} \affiliation{
H.H. Wills Physics Laboratory, University of Bristol, Tyndall Avenue,
 Bristol BS8 1TL, UK}
\affiliation{
Hewlett-Packard Laboratories, Stoke Gifford, Bristol
BS12 6QZ, UK}

\begin{abstract}
The notion of weak measurement provides a formalism for extracting
information from a quantum system in the limit of vanishing
disturbance to its state. Here we extend this formalism to the
measurement of sequences of observables. When these observables do not
commute, we may obtain information about joint properties of a quantum
system that would be forbidden in the usual strong measurement
scenario. As an application, we provide a physically compelling
characterisation of the notion of counterfactual quantum computation.
\end{abstract}


\maketitle
\pagestyle{plain}

\section{Introduction}

Quantum mechanics is still capable of giving us surprises. A good
example is the concept of weak measurement discovered by Aharonov and
his group \cite{AAV88,ABPRT01}, which challenges one of the canonical
dicta of quantum mechanics: that non-commuting observables cannot be
simultaneously measured.

Standard measurements yield the eigenvalues of the measured
observables, but at the same time they significantly disturb the
measured system. In an ideal von Neumann measurement the state of the
system after the measurement becomes an eigenstate of the measured
observable, no matter what the original state of the system was. On
the other hand, by coupling a measuring device to a system weakly it
is possible to read out certain information while limiting the
disturbance to the system. The situation becomes particularly
interesting when one post-selects on a particular outcome of the
experiment. In this case the eigenvalues of the measured observable
are no longer the relevant quantities; rather the measuring device
consistently indicates the {\em weak value} given by the AAV formula
\cite{AAV88,AharonovRohrlich05}:
\be \label{AAV-formula}
A_w=\frac{\bra{\psi_f}A\ket{\psi_i}}{\braket{\psi_f}{\psi_i}}
\ee
where $A$ is the operator whose value is being ascertained,
$\ket{\psi_i}$ is the initial state of the system, and $\ket{\psi_f}$
is the state that is post-selected (e.g. by performing a
measurement). The significance of this formula is that, if we couple a
measuring device whose pointer has position coordinate $q$ to the
system $\cS$, and subsequently measure $q$, then the mean value
$\langle q \rangle$ of the pointer position is given by
\be \label{q-average} \langle q \rangle = g\ Re [A_w], \ee
where $Re$ denotes the real part. This formula requires the initial
pointer wavefunction to be real and of zero mean, but these
assumptions will be relaxed later. The coupling interaction is also
taken to be the standard von Neumann measurement interaction
$H=gAp$. The coupling constant $g$ is assumed to be small, but we can
determine $A_w$ to any desired accuracy if enough repeats of the
experiment are carried out.

The formula (\ref{AAV-formula}) implies that, if the initial state
$\ket{\psi_i}$ is an eigenstate of a measurement operator $A$, then
the weak value post-conditioned on that eigenstate is the same as the
classical (strong) measurement result. When there is a definite
outcome, therefore, strong and weak measurements agree. However, weak
measurement can yield values outside the normal range of measurement
results, eg spins of 100 \cite{Spin100}. It can also give complex
values, whose imaginary part correspond to the pointer momentum. In
fact, the mean of the pointer momentum is given by
\be \label{p-average}
\langle p \rangle = 2gv\ Im [A_w],
\ee
where $Im$ denotes the imaginary part and $v$ is the variance in
the initial pointer momentum.

The fact that one hardly disturbs the system in making weak
measurements means that one can in principle measure different
variables in succession. We follow this idea up in this paper. 

\section{A new paradox}\label{new}

\begin{figure}[hbtp]
\centerline{\epsfig{file=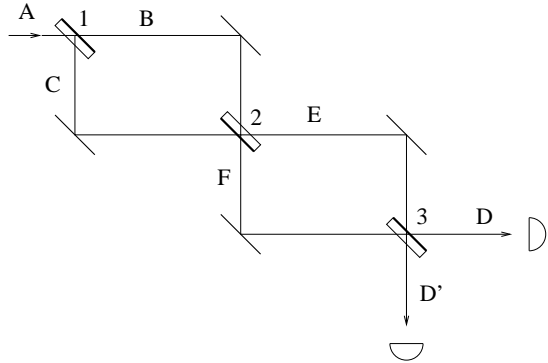,width=0.4\textwidth}}
\caption{The double interferometer: an optical circuit in which a
photon, injected along path $A$, passes through two
interferometers, represented by paths $B$ and $C$ and paths $E$
and $F$. Finally, the photon is post-selected at the detector $D$.
The beam-splitters are shown with their reflecting surface marked
in black. \label{weak2}}
\end{figure}

Weak measurement has proved to be a valuable tool in analysing
paradoxical quantum situations, such as Hardy's paradox
\cite{Hardy92,ABPRT01}.  To illustrate the idea of sequential weak
measurement and its potential applications we first construct a new
quantum paradox.  Consider the {\em double interferometer}, the
optical circuit shown in Figure \ref{weak2}, where a photon passes
through two successive interferometers. This configuration has been
considered previously by Bl\"{a}si and Hardy \cite{BH95} in another
context. Using the labels of the paths shown in the figure, and
denoting the action of the $i$-th beam-splitter by $U_i$, the system
evolves as follows:
\begin{align}
U_1\ket{A}&=(\ket{B}+\ket{C})/\sqrt{2},\label{evolution1} \\
U_2\ket{B}&=(\ket{E}+\ket{F})/\sqrt{2},\ \ U_2\ket{C}=(\ket{E}-\ket{F})/\sqrt{2},\label{evolution2}\\
U_3\ket{E}&=(-\ket{D}+\ket{D'})/\sqrt{2},\ \  U_3\ket{F}=(\ket{D}+\ket{D'})/\sqrt{2}.\label{evolution3}
\end{align}
(The signs here are determined by the fact that reflection on the
silvered outer surface of a beam-splitter gives a phase of $\pi$
whereas transmission or reflection by the inner surface gives zero
phase.)

Suppose now that we select a large number $N$ of successful runs of
our experiment, i.e. those runs where the photon is detected by the
detector $D$.

We can now make the following statements about this situation:

(1) {\it All photons go through path $E$.}

Indeed, equations (\ref{evolution1}) and (\ref{evolution2}) tell us
that if a photon is injected along path A, it must exit the first
interferometer along path $E$. Consequently, if we measure the
observable $P_E$, the projector for path $E$, we find the total number
of photons detected is $N_E=N$ with certainty.

(2) {\it All photons go through path $C$.}

Indeed, the second interferometer is arranged in such a way that any
photon entering along path $B$ will end up at $D'$. Hence, a very
simple calculation shows that if, instead of measuring $N_E$, we
measure $N_C$, the number of photons going along path $C$ in all $N$
runs of the experiment, we will obtain with certainty $N_C=N$.

(3) {\it When photons go through path $C$, a subsequent
   measurement reveals that half of them must go through path $E$ and
   half through path $F$.}

Indeed, if we measure the position of the photons in the first
interferometer and find that all go via $C$, then a subsequent
measurement of $N_E$ and $N_F$ must yield $N/2$ in each case, up to
statistical fluctuations. (In fact this is true regardless of whether or
not all photons end up eventually at $D$).

(4) {\it When photons go through path $E$, a subsequent
   measurement reveals that half of them must have
  come via path $B$ and half via path $C$.}

This last statement is similar to point (3) above.

The above four statements seem to imply a paradoxical situation. On
the one hand, statement (2) tells us, when we pool all the results,
that all $N$ photons go via path $C$; together with statement (3) this
implies that the number of photons that go along path $E$ must be
$N/2$. On the other hand, statement (1) tells us that all $N$ photons
actually go along path $E$! A similar contradiction arises in
connection with the number of photons going along path $C$. On the one
hand, statement (1) tells us that all photons go via $E$; together
with statement (4) this implies that the number of photons that go
along path $C$ must be only $N/2$. On the other hand, statement (2)
tells us that all $N$ photons actually go along path $C$!

The usual way of resolving this paradox is to say that the above
statements refer to measurements that cannot all be made
simultaneously. Indeed, it is true that if we measure $P_E$ we find
it is 1 with certainty, but {\it only} if we do not also measure
$P_C$. If we also measure $P_C$ in the same experiment, then it is no
longer the case that $P_E=1$. Similarly, it is true that $P_C=1$ with
certainty, but {\it only} if we do not also measure $N_E$. If we also
measure $P_E$ in the same experiment, then it is no longer the case
that $P_E=1$. So, we are told, the statements (1)-(4) above have no
simultaneous meaning, for they do not refer to the same
experiment. Hence there is no paradox: In formulating the paradox
presented above we made use of facts that are not all simultaneously
true.

\begin{figure}[hbtp]
\centerline{\epsfig{file=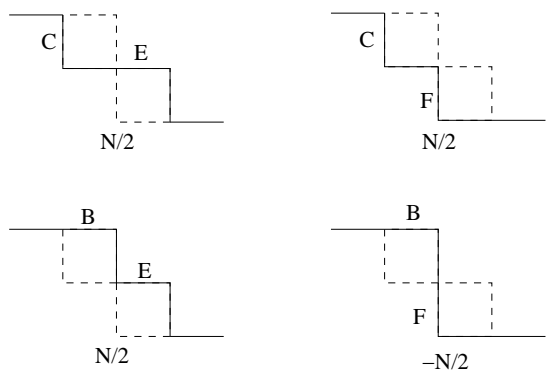,width=0.4\textwidth}}
\caption{Paths through the double interferometer, and the number of
photons that follow the indicated path. Thus for instance
$N_{BE}=N/2$. Note however the curious prediction $N_{BF}=-N/2$.
\label{paths}}
\end{figure}

On the other hand, as is emphasised in \cite{AharonovRohrlich05}, one
should not dismiss such paradoxes too lightly. Indeed it is possible
to make a trade-off: By accepting some imprecision in measuring $P_E$,
$P_C$, etc., we can limit the disturbance these measurements
produce. The way to do this is to weaken the coupling of the measuring
devices to the photons.

Since the disturbance is now small, we can make all the measurements
in the same experiment, and we expect all the statements (1)-(4) to be
true. Hence we expect $N_E=N$, $N_C=N$ and obviously $N_F=0$ and
$N_B=0$. On the other hand, we also expect that $N_{CE}$, and
$N_{CF}$, the total numbers of photons that went along $C$ and
subsequently along $E$ or $F$, respectively, should both be equal to
$N/2$; this is because all the $N$ photons go via $C$ and half of them
should continue along $E$ and half along $F$. Also we expect $N_{CF}$,
the number of photons that went along $C$ and subsequently along $E$,
to be $N_{CE}=N/2$. Similarly we expect that $N_{CE}$ and $N_{BE}$
should both be $N/2$, since all $N$ photons go along $E$ and half of
them must come via $B$ and half via $C$.

While all the above predictions seem reasonable, here is the surprise:
Overall we have only $N$ photons. They could have moved along four
possible trajectories: $BE$, $BF$, $CE$ or $CF$. Since
$N_{BE}+N_{BF}+N_{CE}+N_{CF}=1$ and since $N_{BE}=N_{CE}=N_{CF}=N/2$
it must be the case that $N_{BF}=-N/2$! Furthermore, our prediction
has a remarkable internal consistency. We know that the total number
of photons that go along $F$ must be zero. They can arrive at $F$ in two
ways, either by $BF$ or $CF$. Thus $N_F=N_{BF}+N_{CF}$. As noted above,
$N_{CF}=N/2$, but no photons are supposed to go through $F$. This is due
to the fact that $N_{BF}$ is negative, i.e.  $N_{BF}=-N/2$.

The above predictions seem totally puzzling, no less puzzling than the
original paradox. However, what we have now is not a mere
interpretation that can simply be dismissed. These are now predictions
about the results of real measurements - in particular the weak
measurement of the number of photons that passes along path B and then
along path F. This is a {\it two-time} measurement. 

In general, by ensuring that the measurement interaction is weak, we
can consider {\em sequences} of measurements. Describing such
measurements is the main subject of our paper. In the process, we will
formally derive the strange predictions made above for the double
interferometer, and will discuss the interpretation of weak
measurements. Finally, we apply these ideas to counterfactual
computation, which is a catch-all for numerous counterfactual
phenomena including, for example, interaction-free measurement
\cite{ElitzurVaidman93}.

\section{Sequential weak measurements}\label{section:seq}

The situation we shall consider is where a system $\cS$ evolves
unitarily from an initial state $\ket{\psi_i}$ to a final
post-selected measurement outcome $\bra{\psi_f}$. At various points,
observables may be measured weakly. Here we consider the scenario
where there is a single copy of the system, with the measuring device
weakly coupled to it. Generally, reliable information will only be
obtained after many repeats of the given experiment.

In the simplest case where there is just one observable, $A$ say, we
assume the evolution from $\ket{\psi_i}$ to the point where $A$ is
measured is given by $U$, and from this point to the post-selection
the evolution is given by $V$. Then we can rewrite (\ref{AAV-formula})
as:
\be \label{fullAAV}
A_w=\frac{\bra{\psi_f}VAU\ket{\psi_i}}{\bra{\psi_f}VU\ket{\psi_i}},
\ee
and the mean of the pointer is given by (\ref{q-average}) as before.

Consider next the case of two observables, $A_1$ and $A_2$, measured
at different times on a system $\cS$. We assume the system evolves
under $U$ from $\ket{\psi_i}$ to the point where $A_1$ is measured,
then under $V$ to the point where $A_2$ is measured, and finally under
$W$ to $\ket{\psi_f}$.  Our strategy is to use two measuring devices
for measuring $A_1$ and $A_2$. Let the positions of their pointers be
denoted by $q_1$ nd $q_2$, respectively. We couple them to the system
at successive times, measure $q_1$ and $q_2$, and then take the
product $q_1q_2$.

We begin, therefore, with the weak coupling of system and pointers,
with the usual von Neumann-type Hamiltonians for measuring $A_1$ and
$A_2$. The state of system and pointers after this coupling is:
\be \label{ABinitial}
\Psi_{\cS\cM_1\cM_2}=e^{-ig p_2 A_2}Ve^{-ig p_1 A_1} U\ket{\psi_i}_{\cS}
\phi(q_1)\phi(q_2),
\ee
where $p_1$ and $p_2$ are the two pointer momenta (the label
$\cS$ refers to the system and $\cM_1$, $\cM_2$ to the pointers). Here
$\phi(q)$ is the initial pointer distribution, and we have assumed,
for simplicity, that the two pointers have identical initial
distributions and equal coupling constants $g$. Post-selecting on
$\bra{\psi_f}$ gives the state of the pointers as
\be \label{ABpointer-only}
\Psi_{\cM_1\cM_2}=\bra{\psi_f}We^{-ig p_2 A_2}Ve^{-ig p_1 A_1} U\ket{\psi_i} \phi(q_1)\phi(q_2).
\ee
As $g$ is small, we can approximate the state as:
\be \label{ABexpansion}
\Psi_{\cM_1\cM_2}=\bra{\psi_f}\left(W(1 -igp_2A_2 -\frac{g}{2}^2p_2^2A_2^2 + \ldots)V
(1 -igp_1A_1 -\frac{g}{2}^2p_1^2A_1^2 + \ldots)U\right)\ket{\psi_i} \phi(q_1)\phi(q_2).
\ee
Putting $p=-i\partial/\partial q$, we get
\bea \label{ABstate} \Psi_{\cM_1\cM_2} &=& F \
\mbox{\big[}\phi(q_1)\phi(q_2)-g(A_1)_w\phi^\prime(q_1)\phi(q_2)-g(A_2)_w\phi(q_1)\phi^\prime(q_2)
+\frac{g^2}{2}(A_1^2)_w\phi^{\prime\prime}(q_1)\phi(q_2)\\\nonumber
&+&\frac{g^2}{2}(A_2^2)_w\phi(q_1)\phi^{\prime\prime}(q_2)+g^2(A_2,A_1)_w\phi^\prime(q_1)\phi^\prime(q_2)+O(g^3)\mbox{\big]}
\nonumber \eea
where $F=\bra{\psi_f}WVU\ket{\psi_i}$,
$(A_1)_w=\bra{\psi_f}WVA_1U\ket{\psi_i}/F$,
$(A_1^2)_w=\bra{\psi_f}WVA_1^2U\ket{\psi_i}/F$,
$(A_2)_w=\bra{\psi_f}WA_2VU\ket{\psi_i}/F$,
$(A_2^2)_w=\bra{\psi_f}WA_2^2VU\ket{\psi_i}/F$ and $(A_2,A_1)_w$ is defined by
\be \label{ABweakvalue}
(A_2,A_1)_w=\frac{\bra{\psi_f}WA_2VA_1U\ket{\psi_i}}{\bra{\psi_f}WVU\ket{\psi_i}}.
\ee

Following measurement of $q_1$ and $q_2$, the expected value of their
product is given by
\be \label{product-formula} \langle q_1q_2 \rangle = \frac{\int
q_1q_2|\Psi_{\cM_1\cM_2}|^2 dq}{\int |\Psi_{\cM_1\cM_2}|^2 dq}.  \ee
For simplicity, let us make the following assumption (we will discuss
the general case later):\\
\newline
{\bf Assumption A}: {\em The initial pointer distribution $\phi$ is
real-valued, and its mean is zero, i.e.  $\int q \phi^2(q)dq=0$}.
\newline
\newline
We also assume, without loss of generality, that $\phi$ is normalised
so that $\int \phi^2=1$. With these assumptions, all the terms in
(\ref{product-formula}) of order 0 and 1 in $g$ vanish, and we are
left with
\begin{align}\label{qAqB}
\langle q_1q_2 \rangle=g^2 \mbox{\big [}
(A_2,A_1)_w+\overline{(A_2,A_1)}_w+\overline{(A_1)}_w(A_2)_w+(A_1)_w\overline{(A_2)}_w \mbox{\big ]} \left(\int q\phi(q)\phi^\prime(q)dq\right)^2,
\end{align}
where bars denote complex conjugates. Integration by parts implies
$\int q \phi(q) \phi^\prime(q)dq=-\frac{1}{2}$, so we get the final
result
\be \label{ABmean} \langle q_1q_2 \rangle=\frac{g^2}{2}\ Re \left[
(A_2,A_1)_w+(A_1)_w\overline{(A_2)}_w \right].  \ee
Here $(A_2,A_1)_w$ is the sequential weak value given by
(\ref{ABweakvalue}); note the reverse order of operators, to fit with
the convention of operating on the left.

\section{The sequential weak value}\label{sequential-weak-value}

In the section above we considered two measurements -- a measurement
of $A_1$ at time $t_1$ and of $A_2$ at $t_2$ -- and we looked at the
product of the outcomes $q_1q_2$ in the limit when the coupling of the
measuring devices with the measured system was weak. This procedure
was motivated by our example of the double interferometer: we wanted
to check whether the photon followed a given path, say the path that
goes along $C$ in the first interferometer and then along $E$ in the
second interferometer. In that case the variables of interest are
$P_C$, the projector on path C and $P_E$, the projector on path
$E$. When the photon follows this path, the value of the
{\it product} of these projectors is $1$ while in all other situations
the product is $0$.  We wanted to see what the behavior of the photon
was when the measurements did not disturb it significantly.

Since $q_1$ measures $A_1$ and $q_2$ measures $A_2$, it seems obvious
that the quantity that represents the product of the two observables
is $\langle q_1q_2 \rangle$ given in (\ref{qAqB}) above. However, the
situation is more subtle, as we show below.

Consider the simpler case of two {\it commuting} operators $A_1$ and
$A_2$, and suppose we are interested in the value of the product
$A_2A_1$ at some time $t$. (Note that we are now talking about operators at
one given time, not at two different times.) We can measure this
product in two different ways. First, we can measure the product
directly, by coupling a measuring device directly to the product via
the interaction Hamiltonian $H=gpA_2A_1$. When we make the coupling
weaker, we find that the pointer indicates the value 
\begin{align} 
\langle
q\rangle=gRe(A_1A_2)_w =gRe
\frac{\bra{\psi_f}A_2A_1\ket{\psi_i}}{\braket{\psi_f}{\psi_i}}. 
\end{align}
This is straightforward: it is simply the weak value of the operator
$A_2A_1$. On the other hand, we could attempt to measure the product
in the same way that we measured the sequential product. That is, we
can use two measuring devices with pointer position variables $q_1$
and $q_2$, couple the first measuring device to $A_1$ and the second
to $A_2$, and then look at the product $q_1q_2$. The latter method was
proposed by Resch and Steinberg \cite{RS04} for the
simultaneous measurement of two operators. They showed that in this case
\begin{align} \label{RSmean} 
\langle q_1q_2 \rangle=\frac{g^2}{2}\ Re\
\left[(A_1A_2)_w+(A_1)_w\overline{(A_2)}_w \right]. 
\end{align} 
We see that the value indicated by $ \langle q_1q_2 \rangle$ is {\it
not} equal to the weak value of the product, but contains a
supplementary term, $Re(A_1)_w\overline{(A_2)}_w $. In other words,
although we expected the two methods to be equivalent, it is not the
case. To obtain the true weak value of the product we must subtract
this second term. This second term is an artifact of the method of
using two separate measuring devices rather than coupling one
measuring device directly to the product operator.
 
In the case of sequential measurement there is no product operator to
start with, for we are interested in the product of the values of
operators at two different times. Hence the first method, of coupling
directly to the product operator, makes no sense, and we must use two
independent couplings. In order to obtain the quantity of interest,
i.e.  the quantity that is relevant to situations such as the double
interferometer of Section \ref{new}, we must subtract the term
$Re(A_1)_w\overline{(A_2)}_w $ from (\ref{ABmean}). We thus conclude
that the quantity of interest is the sequential weak value given in
(\ref{ABweakvalue}).

\section{General sequential weak measurement}\label{general}

Sequential weak measurement can be easily extended to $n$ measurements
of Hermitian operators $A_i$ with intervening unitary evolution steps
$U_i$. The weak values are given by
\begin{align} \label{sequential-value}
(A_n, \ldots ,A_1)_w=\frac{\bra{\psi_f}U_{n+1}A_nU_n \ldots A_1U_1\ket{\psi_i}}{\bra{\psi_f}U_{n+1}U_n \ldots U_1\ket{\psi_i}},
\end{align}
and the expected values $\langle q_1q_2 \ldots q_n \rangle$ can be
expressed in terms of these weak values. For example, with
Assumption A
\be \label{123mean}
\langle q_1q_2q_3 \rangle=\frac{g^3}{4} \ Re \left[
(A_3,A_2,A_1)_w+(A_2,A_1)_w\overline{(A_3)}_w+(A_3,A_1)_w\overline{(A_2)}_w+(A_3,A_2)_w\overline{(A_1)}_w \right],
\ee
and the case of general $n$ is given in the Appendix. Similarly, we
can express expected values for products of momenta in terms of the
weak values (see Appendix). For instance
\begin{align} \label{special}
\langle p_1p_2 \rangle = 2(gv)^2Re \left[-(A_2,A_1)_w+(A_1)_w\overline{(A_2)}_w \right].
\end{align}
Mixed products of positions and momenta give similar formulae. For instance
\begin{align} \label{mixed}
\langle q_1p_2 \rangle=-g^2v \ Im
\left[(A_2,A_1)_w+\overline{(A_1)}_w(A_2)_w \right].
\end{align}

The foregoing examples illustrate a general pattern, which is that
expectations of products of $p$'s and $q$'s depend on the real part of
sequential weak values if there is an even number of $p$'s in the
product and on the imaginary part if there is an odd number of $p$'s.

The sequential weak values satisfy the following rules:

\vspace{0.2cm}
1) {\bf Linearity in each variable separately:}
\[
(A_n, \ldots , A_i , \ldots , A_1)_w+(A_n, \ldots , A_i^\prime , \ldots , A_1)_w=(A_n, \ldots , (A_i+A_i^\prime) , \ldots , A_1)_w,
\]
for any $1 \le i \le n$.

2) {\bf Agreement with strong measurement:}

Suppose that, with preselection by $\ket{\psi_i}$ and post-selection
by $\ket{\psi_f}$, strong measurements of $A_1$, $A_2$, $\ldots$ ,
$A_n$ always give the same outcomes $a_1, a_2, \ldots, a_n$; then
$(A_n \ldots A_1)_w=a_1a_2 \ldots a_n$.

3) {\bf Marginals:} If $I$ is the identity operator at location $i$:
\[
(A_n, \ldots A_{i+1}, A_{i-1}, \ldots , A_1)_w=\sum_i(A_n, \ldots
A_{i+1}, I, A_{i-1}, \ldots , A_1)_w.
\]
\vspace{0.1cm}

We can illustrate some of these rules with the double interferometer
experiment (figure \ref{weak2}). The measurements we consider are
projectors that detect the presence of a photon on various edges; for
instance, the projector $P_B$ indicates whether a photon is present on
the edge $B$. For simplicity we write $B_w$ for the weak value
$(P_B)_w$, etc., and we use the same convention for sequential weak
values. Then using (\ref{fullAAV}) we find $C_w=1$, $B_w=0$, $E_w=1$
and $F_w=0$. Using (\ref{ABweakvalue}) we find $(E,B)_w=1/2$,
$(F,B)_w=-1/2$, $(E,C)_w=1/2$ and $(F,C)_w=1/2$. Since $P_E+P_F=I$,
rule 1) implies $(E,B)_w+(F,B)_w=(I,B)_w$, and then rule 3) implies
$(I,B)=B_w$. Thus we expect $(E,B)_w+(F,B)_w=B_w$, which holds if we
substitute the values above. Similarly $(E,C)_w+(F,C)_w=1/2+1/2=C_w$,
and so on. As for rule 2), we have seen (Section \ref{new}) that
strong measurement of $P_C$ and $P_E$ yields 1, so we expect the weak
values to be the same, as is the case.

\begin{figure}[hbtp]
\centerline{\epsfig{file=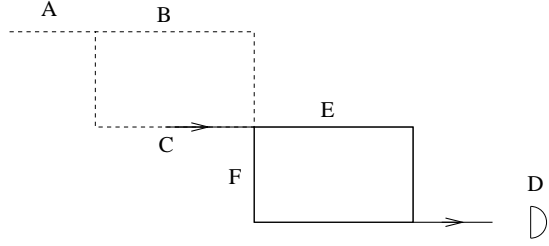,width=0.4\textwidth}}
\caption{The double interferometer restricted to its second
interferometer. According to (\ref{ratio-rule}), the ratio of the weak
values $E_w/F_w$ in the second interferometer, with photons injected
along $C$, is the same as the ratio of the sequential weak values
$(E,C)_w/(F,C)_w$ in the double interferometer with photons injected
along $A$.
\label{small}}
\end{figure}

There is a further rule that applies when one of the operators being
measured is a projector. We illustrate it with the double
interferometer. We can write
\begin{align}\label{ratio-rule}
\frac{(E,C)_w}{(F,C)_w}=\frac{\bra{D}U_3P_EU_2\ket{C}\
\bra{C}U_1\ket{A}\ }{\bra{D}U_3P_FU_2\ket{C}\ \bra{C}U_1\ket{A}\
}=\frac{\bra{D}U_3P_EU_2\ket{C}}{\bra{D}U_3P_FU_2\ket{C}}=\frac{E_w}{F_w}.
\end{align}
Here $E_w$ and $F_w$ in the final ratio are calculated assuming that
$\ket{\psi_i}=\ket{C}$, in other words, as though we were calculating
weak values for the second interferometer treated separately from the
rest of the system, with initial state $\ket{C}$ and post-selection by
$\ket{D}$ (Figure \ref{small}). If we only knew the
single-measurement weak values $E_w$, $F_w$ and $C_w$, we could
calculate $(E,C)_w$ and $(F,C)_w$ using this rule and the relationship
$(E,C)_w+(F,C)_w=C_w$ derived above.

\section{The meaning of weak values}\label{meaning:seq}

Consider some experiment in which we inject some kind of particle and
weakly measure the projector onto some location $X$. Suppose we
collect some large number $N$ of runs of the experiment that satisfy
the post-selection criterion. We interpret the fact that the projector
at $X$ has weak value $X_w$ to mean that, for any appropriate physical
property we test, due for instance to the charge, gravitational field,
etc. of the particle, it is as though $NX_w$ particles (up to a
binomial distribution error) passed along $X$. Thus in the double
interferometer experiment we expect all physical tests to give
outcomes appropriate to there being, in all $N$ runs of the
experiment, a total of $N_E=NE_w=N$ photons passing along $E$,
$N_{CE}=N/2$ photons passing along $C$ then $E$, and so on.

Can we justify the foregoing interpretation of weak values? For weak
measurements of a single operator, there is a body of work showing
that weak values, even when they lie in an unexpected range, can be
treated as though they were the actual values in the underlying
physical theory and will then yield correct predictions. Examples of
this include weakly measured negative kinetic energies when a particle
is in a classically forbidden region \cite{RAPV95}, and weakly
measured faster-than-light velocities that are associated with
Cerenkov radiation \cite{RA02}. If a measure is entirely consistent
with physics in this fashion, then we are entitled to say that it is
telling us a true physical fact. For sequential weak values, we can
make a similar argument. The physical meaning of sequential weak
values needs to be explored in many physical situations to give the
kind of justification that single weak values enjoy. However, the
internal consistency is already clear from the double interferometer
example, and, more generally, from the rules in Section \ref{general}.

\section{Broadening the concept: weak interactions} \label{broad}

So far, we have considered ideal weak measurements, in which the
pointer distribution is real and has zero mean (Assumption A). If
we drop these assumptions, we find in place of (\ref{q-average})
that
\begin{align}\label{complex-version}
\langle q \rangle=\mu+g(Re[A_w]+Im[A_w]y),
\end{align}
where $y=\int \bar \phi(pq+qp)\phi dq - 2\mu \nu$, with $\mu=\int \bar \phi
q \phi dq$, $\nu=\int \bar \phi p \phi dq$.

The expectation $\langle r_1r_2 \ldots r_n\rangle$ for a general
initial pointer distribution, where each $r_i$ is either $q_i$ or
$p_i$, is a very complicated expression, but, so far as the system
goes, depends only on the real and complex parts of sequential weak
values up to $(A_n,\ldots A_1)_w$. Thus we can write
\begin{align}\label{poly-form}
\langle r_1r_2 \ldots r_n \rangle=\Phi(Re(A_n,\ldots A_1)_w,Im(A_n,\ldots A_1)_w, \ldots ,Re(A_n)_w,Im(A_n)_w, \ldots , Re(A_1)_w,Im(A_1)_w),
\end{align}
for some polynomial function $\Phi$. The coefficients in $\Phi$
are themselves polynomials in expectations $\int \bar f
\gamma(p_i,q_i)f dq$ for polynomials $\gamma$, as we see in the
case of equation (\ref{complex-version}), where $y$ has this form.

In the next section, we shall want to consider the most general
possible type of {\em weak interaction} which allows any sort of
(suitably weak) coupling between the system and an ancilla followed by
any further evolution or measurement of the ancilla alone (the pointer
in our previous discussion and its von Neumann measurement interaction
$gpA$ will be a special case of such an ancilla and weak
interaction). Our notion of general weak interaction is the following:
Consider the system and ancilla initially in product state
$\ket{\psi_i}\ket{\xi}$. Let $H_{\rm S,anc}$ be any Hamiltonian of the
joint system, and $g$ a coupling constant. For a single interaction
event, and to first order in $g$, the state becomes
\begin{align}\label{eq1}
(I-igH_{\rm S,anc})\ket{\psi}\ket{\xi}.
\end{align}
Any joint Hamiltonian may be expressed as a sum of products of
individual Hamiltonians
\begin{align}
H_{\rm S,anc}= \sum_k H^k_{\rm S} \otimes H^k_{\rm anc}.
\end{align}
Post-selecting the system state in equation (\ref{eq1}) with
$\ket{\psi_f}$ gives
\begin{align}
\Psi_{\rm anc}=\braket{\psi_f}{\psi_i}[I_{\rm anc}-ig\sum_k(H^k_{\rm
S})_wH^k_{\rm anc}]\ket{\xi};
\end{align}
So the system Hamiltonians $H^k_{\rm S}$ have been effectively
replaced by their weak values $(H^k_{\rm S})_w$. The important point
here is that all subsequent manipulations of the ancilla will depend
on the pre- and post-selected system only through {\em weak values} of
suitably chosen observables. A similar result clearly holds for any
sequential weak interactions and suitably associated sequential weak
values, and also for terms of any higher order in $g$.

As a simple illustrative example, suppose that the ancilla is the
pointer system of a von Neumann measurement interaction with
Assumption A in force, and that this same pointer is weakly coupled
twice for the sequential measurement of both $A_1$ and $A_2$.  If this
pointer has position $q$ and momentum $p$, the pointer state after
post-selection is
\begin{align}
\Psi_{\cM}&=\bra{\psi_f} \left(U_3e^{-igpA_2}U_2e^{-igpA_1}U_1\right) \ket{\psi_i} \phi(q),
\end{align}
yielding
\[
\langle q \rangle=g\ Re \ \left[(A_1)_w+(A_2)_w \right].
\]
The effect in this instance is therefore the same as adding the
individual post-measurement results, and it depends on the system
only through associated weak values.

\section{Counterfactuality and weak measurement}

Counterfactual computation \cite{Jozsa98,MitJozsa} provides a general
framework for looking at counterfactual phenomena, including
interaction-free measurement as a special case. We consider arbitrary
protocols, at various points of which a quantum computer can be
inserted. The computer has a switch qubit (with $\ket{0}$=off and
$\ket{1}$=on) and an output qubit. A special case of this formalism is
where the protocol is represented by an optical circuit, and a computer
insertion means that the computer (or a copy of it) is placed in some
path of the circuit and is switched on by a photon passing along that
path.

We assume that the computer is programmed ready to perform a
computational task with answer $0$ or $1$ which will be written into
the output qubit if the switch is turned on. In addition to the switch
and output qubits, the protocol will in general have additional
qubits, and will involve some measurements. We say that an outcome of
these measurements {\em determines the computer output} if that
outcome only occurs when the computer output has a specific value,
$\ket{0}$ or $\ket{1}$.  Such an outcome is said to be counterfactual
if its occurrence also implies that the computer was never switched
on, i.e. its switch was never set to $\ket{1}$, during the protocol.

To make this precise, note first that one can always produce an
equivalent protocol in which the state is entangled with extra qubits
and the measurement deferred to the end of the protocol. Thus the
protocol can be assumed to consist of a period of unitary evolution
followed by a measurement, which can be assumed (again by adding extra
qubits) to be a projective measurement. Let $\ket{\psi_i}$ be the
initial state of the protocol, and let $\ket{\psi_f}$ be a measurement
outcome that determines some specific computer output, in the sense
defined above. Suppose the computer is inserted $n$ times. Let $\cF$
(for ``oFf'') denote the projection $\ketbra{0}$ onto the off value of
the computer switch and $\cN$ (for ``oN'') denote the complementary
projector $\ketbra{1}$, and let $\xi$ be one of the $2^n$ possible
strings of $\cF$'s or $\cN$'s of length $n$; we call this a {\em
history}. Let $U_i$ denote the unitary evolution in the protocol
between the $(i-1)$th and $i$th insertions of the computer.

\begin{definition}[Counterfactuality by histories \cite{MitJozsa}]
\label{histories}
The measurement outcome $\ket{\psi_f}$ is a {\em counterfactual
outcome} if

1) $\ket{\psi_f}$ determines the computer output.

2) The amplitude of any history $\xi$ containing an $\cN$ vanishes. In
other words, for all histories $\xi$ other than the all-$\cF$ history,
$\bra{\psi_f}U_{n+1}\xi_nU_n \ldots U_2\xi_1 U_1\ket{\psi_i}=0$.
\end{definition}

One may question whether this is the ``correct'' definition of a
notion of counterfactual computation or whether alternative
definitions might be convincingly plausible. Condition 1) is
uncontroversial but condition 2) might seem less immediately
compelling. It is evidently equivalent to obtaining a null result if
we carry out a strong non-demolition measurement of {\cal N} at each
computer insertion. However the disturbance that such a measurement
causes might lead one to question the suitability of this
condition. Indeed recently Hosten et al. \cite{Hosten06} proposed an
alternative definition of counterfactual computation that violates
condition 2) of definition VIII.1 and sparked a controversy
\cite{MJ06} over the relative merits and validity of the two
notions. We will now develop some alternative characterisations of our
definition VIII.1 in terms of {\em weak} measurements, thereby
addressing the disturbance issue. We will argue that these new
characterisations considerably strengthen the credibility of the
original definition as the ``correct'' one.

Let us therefore consider carrying out a weak measurement of $\cN$ at
each insertion. A non-zero weak value implies that there is a
detectable physical effect that can only occur if the computer is
switched on. Vaidman's treatment of the three-box paradox
\cite{Vaidman06} gives a good example of this reasoning.

\begin{figure}
\centerline{\epsfig{file=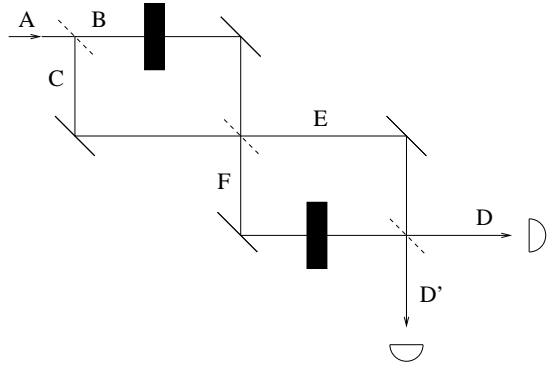,width=0.4\textwidth}}
\caption{The double interferometer of Figure \ref{weak2} treated as a
protocol with computer insertions (black rectangles) in paths $B$ and
$F$. If a photon passes down either of these paths, the computer runs.
\label{weak3}}
\end{figure}

Our two-interferometer example shows that it does not suffice to
consider the individual weak values at each insertion. For suppose the
computer is inserted in paths $B$ and $F$, as shown in Figure
\ref{weak3}. Then we have seen that the weak values $B_w$ and $F_w$
are zero, yet the sequential weak value $(F,B)_w$ is non-zero. The
non-vanishing of the sequential weak value implies that a photon
passes along {\em both} path $B$ and $F$, since there is a physical
effect that causes correlated deflections of pointers at both sites.

There is a subtlety here, because it could be argued that, because
sequential pairwise weak measurements give second-order effects in $g$
(see (\ref{ABmean})), we might detect a departure from zero in the weak
measurements for each operator individually, i.e. in the deflections
of the pointers at $B$ and $F$, if we looked at second or higher order
terms in $g$. However, if $A$ is any projector and $A_w=0$, then the
von Neumann interaction $e^{-igpA}$ reduces to $Ae^{-igp}+I-A$, which
is the identity to all orders in $g$ in the weak measurement
calculation. Thus we truly need to carry out the sequential weak
measurement here to identify the physical effect due to the photon.

In general, we need to consider all possible sequential weak
measurements to obtain an adequate test of counterfactuality. This is
why we must use weak rather than strong measurements. As we have seen
in Section \ref{sequential-weak-value}, there is no strong measurement
corresponding to sequential weak measurements.

We therefore propose the following:

\begin{definition}[Counterfactuality by weak values]\label{weak-values}
The measurement outcome $\ket{\psi_f}$ is a {\em counterfactual
outcome} if

1) $\ket{\psi_f}$ determines the computer output.

2) $(\cN_{i_k}, \cN_{i_{k-1}}, \ldots \cN_{i_1})_w=0$,
for any $1 \le i_1 < i_2 < \ldots < i_k \le n$, where $n$ is the number of insertions of the computer.
\end{definition}

By (\ref{sequential-value}), conditions 2) for \ref{histories} and
\ref{weak-values} are equivalent, using the fact that $\cF+\cN=1$
together with the linearity and marginal rules. For instance, with two
insertions of the computer, condition 2) of Definition \ref{histories}
amounts to $(\cN_1,\cN_2)_w=0$, $(\cF_1,\cN_2)_w=0$ and
$(\cN_1,\cF_2)_w=0$, and these imply $(\cN_1)_w=0$, $(\cN_2)_w=0$ and
$(\cN_1,\cN_2)_w=0$, which constitute condition 2) for Definition
\ref{weak-values}.

We can try to strengthen the requirements for counterfactuality by
demanding that a zero response is obtained for any conceivable
weak interaction, in the sense of the preceding section. In our
present application we must further restrict the weak interaction
to take place only if the switch has the property of being "on",
i.e. the interaction Hamiltonian must have the form $(\cN \otimes
I_{\rm anc})H_{\rm s, anc} (\cN \otimes I_{\rm anc})$. We say that
such an interaction is a weak interaction involving the projector
$\cN$. Since $\cN$ is a one-dimensional projector, this implies
that the interaction Hamiltonian has the form $\cN \otimes H_{\rm
anc}$. In a more general scenario the projector $\tilde \cN$ for
counterfactuality (analogous to the switch being "on") may have
rank larger than 1 and then the interaction Hamiltonian may have
the more general form $(\tilde \cN \otimes I_{\rm anc})M_{\rm s,
anc} (\tilde \cN \otimes I_{\rm anc})$ for any Hermitian $M$. For
example, the switch may be a photon with both path and
polarisation properties. Then a weak interaction restricted to its
presence on a path would correspond to a two-dimensional projector
on its polarisation state-space associated to that path.

\begin{definition}[Counterfactuality by general weak interactions]\label{universal}
The measurement outcome $\ket{\psi_f}$ is a {\em counterfactual
outcome} if

1) $\ket{\psi_f}$ determines the computer output.

2) Any possible weak interaction involving the projections $\cN_1,
\ldots \cN_n$ yields a null result.
\end{definition}

By a null result, we mean the same result that would be obtained
for $g=0$. It is not difficult to show that this apparently much
broader concept is in fact equivalent to Definition
\ref{weak-values}. In one direction, we know from the last section
that any expectation depends only on the sequential weak values,
involving the projectors $\cN_i$, so when these weak values vanish
we obtain a null result. In the other direction, we have only to
show that we can choose particular weak interactions whose null
results will imply the vanishing of all sequential weak values.
However, if we first obtain a null value of $\langle q_i \rangle$
and $\langle p_i \rangle$ for the standard von Neumann measurement
weak interaction for every $i$, then we know by (\ref{q-average})
and (\ref{p-average}) that both real and imaginary parts of all
the weak values $(\cN_i)_w$ are zero. Then by obtaining null
values of $\langle q_iq_j \rangle$ and $\langle p_iq_j \rangle$
for all $i < j$, we infer from (\ref{ABmean}) and (\ref{mixed})
that the real and imaginary parts of all $(\cN_j,\cN_i)_w$ are
zero. We continue this way, using the fact that expectations of
products of $p$'s and $q$'s with an even number of $p$'s depend on
the real part of sequential weak values, whereas those with an odd
number of $p$'s depend on their imaginary parts (see Appendix).

We have therefore proved:
\begin{theorem}
All three definitions, \ref{histories}, \ref{weak-values} and
\ref{universal}, are equivalent.
\end{theorem}

\section{Discussion}

Sequential weak values are a natural generalisation of the weak value
of a single measurement operator \cite{AAV88}. Resch and Steinberg's
simultaneous measurement of two operators \cite{RS04} gives the same
result in the special case where these operators commute, but it
does not address the case where we have a succession of measurements
with unitary evolution between them.

One can argue that both single and sequential weak measurements tell
us what the physical situation is. In the double interferometer, for
instance, $C_w=1$ really means that all the photons go via $C$, and
$(E,C)_w=1/2$ really means that approximately half the photons go via
$C$ followed by $E$. This is of course a matter of interpretation, and
may be disputed; but at least it seems to be true that weak values can
be fitted into the framework of physics without contradiction, and
give illuminating explanations of many phenomena.

Our application of weak measurement to counterfactuals does not depend
on the foregoing interpretation. The most straightforward part of our
claim is that, if a weakly coupled measuring device indicates a
displacement of pointers in some region of an apparatus, then one
cannot claim that the state of the system was unaltered in that
region; for example, in the case of an optical device, such a shift
would indicate that a photon was present. The importance of sequential
weak measurements in this context is illustrated by the double
interferometer (Figure \ref{weak2}). If two pointers are coupled to
the paths $B$ and $F$ in this apparatus, each pointer individually
will show no displacement on average after many runs of the
experiment. However, the product of the positions of the pointers will
show a shift. Thus the photon reveals its presence only when
information from both pointers is suitably combined.

The other part of our claim about counterfactuals can be summed up by
what we might call the principle of weak detectability:

{\em An event that cannot be detected by any possible weak
interaction does not take place.}

This means that we learn a fact $X$ about an event
counterfactually from a certain experiment if (1) the outcome of
the experiment implies $X$, and (2) no possible weak interaction
can detect the occurrence of this event during the experiment. It
seems as though part (2) might be hard to confirm, because there
is a great variety of possible weak interactions.  However, this
condition proves to be equivalent to the vanishing of all
sequential weak values associated to the event in question, and
this will often be much easier to check.

Finally, we mention the striking fact that sequential weak values are
formally closely related to amplitudes. Consider the case where we
measure $n$ projectors $P_{X_1}, \ldots P_{X_n}$ that define a path
$\pi_x$ between the initial and post-selected states $\ket{\psi_i}$
and $\ket{\psi_f}$, respectively. We can write
\begin{align}
(P_{X_n}, \ldots , P_{X_1})_w= \frac{\bra{\psi_f}U_{n+1}\ket{X_n}\
\bra{X_n}U_n\ket{X_{n-1}} \ \ldots
\bra{X_1}U_1\ket{\psi_i}}{\bra{\psi_f}U_n \ldots U_1\ket{\psi_i}}=\frac{{\rm Amplitude} (\pi_x)}{\sum_i \mbox{Amplitude}(\pi_i)},
\end{align}
where $\pi_i$ runs over all paths between $\ket{\psi_i}$ and
$\ket{\psi_f}$. Nonetheless, weak values are like measurement results
rather than amplitudes! This way of looking at sequential weak values
suggests a close connection with path integrals that remains
to be explored.

\section*{ACKNOWLEDGEMENTS}
We thank L. Vaidman for helpful comments on an earlier version. GM
acknowledges support from the project PROSECCO~(IST-2001-39227) of the
IST-FET programme of the EC. RJ and SP are supported by the EPSRC
QIPIRC and EC networks QAP and QICS, and SP also acknowledges support
from the EPSRC grant GR/527405/01.

\appendix
\section{Calculation of general correlations}

With Assumption A, we show here that the general version of
(\ref{ABmean}) is
\begin{align} \label{sequential-product}
\langle q_1q_2 \ldots q_n \rangle=\frac{g^n}{2^{n-1}}\ Re \sum_{r
\ge s} \sum_{\bf i,j} (A_{i_r}, \ldots ,A_{i_1})_w\overline{(A_{j_s}, \ldots ,A_{j_1})}_w.
\end{align}
where the weak values in this formula are given by \ref{sequential-value}.
In (\ref{sequential-product}) the sum is over all ordered indices
${\bf i}=(i_1, \ldots i_r)$ with $i_p < i_{p+1}$ for $1 \le p \le
r-1$, and ordered indices ${\bf j}=(j_1, \ldots j_s)$ that make up the
complement of ${\bf i}$ in the set of integers from $1$ to $n$,
i.e. that satisfy $(i_1, \ldots i_r)\cup (j_1, \ldots j_s)= (1,2,
\ldots n)$ and $(i_1, \ldots i_r)\cap (j_1, \ldots j_s)=
\emptyset$. We include the empty set $\emptyset$ as a possible set of
indices. In order not to count indices twice, we require $r \ge s$,
and when $r=s$ we require $i_1=1$.

For instance, with $n=2$, the possible indices are ${\bf i} =(1,2)$,
${\bf j} =\emptyset$; ${\bf i} =(1)$, ${\bf j} =(2)$, which yields
\be \label{12mean}
\langle q_1q_2 \rangle=\frac{g^2}{2} \ Re \left[
(A_2,A_1)_w+(A_1)_w\overline{(A_2)}_w \right].
\ee
This is just equation (\ref{ABmean}). For $n=3$ we have ${\bf i}
=(1,2,3)$, ${\bf j} =\emptyset$; ${\bf i} =(1,2)$, ${\bf j} =(3)$;
${\bf i} =(1,3)$, ${\bf j} =(2)$; ${\bf i} =(2,3)$, ${\bf j} =(1)$,
giving (\ref{123mean}). Equation (\ref{sequential-product}) is proved
in the same way as (\ref{ABmean}), the state of the $n$ pointers after
post-selection being:
\bea \label{q-expansion} \Psi_{\cM_1 \ldots
\cM_n}&=&\bra{\psi_f}\left(U_{n+1}e^{-igp_nA_n}U_n \ldots
U_2e^{-igp_1A_1}U_1\right)\ket{\psi_i} \phi(q_1) \ldots \phi(q_n),
\\\nonumber
&=&\bra{\psi_f}\left(U_{n+1}\left(\phi(q_n)-gA_n\phi^\prime(q_n)+ \ldots
\right)U_n \ldots U_2\left(1-gA_1\phi^\prime(q_1) + \ldots
\right)U_1\right)\ket{\psi_i}, \\\nonumber &=&
\bra{\psi_f}U_{n+1}U_n \ldots U_1\ket{\psi_i} \ \left(1 + g\sum_i
\frac{\phi^\prime(q_i)}{\phi(q_i)}(A_i)_w + g^2\sum_{i<j}
\frac{\phi^\prime(q_i)\phi^\prime(q_j)}{\phi(q_i)\phi(q_j)}(A_j,A_i)_w +
\ldots \right)\phi(q_1) \ldots \phi(q_n).\nonumber \eea
Assumption A implies that only the terms in $q_1q_2 \ldots q_n$ in
$|\Psi_{\cM_1 \ldots \cM_n}|^2$ need to be taken into account in
calculating
\[
\langle q_1q_2 \ldots q_n \rangle = \frac{\int q_1q_2 \ldots q_n
|\Psi_{\cM_1 \ldots \cM_n}|^2 dq_1 \ldots dq_n}{\int |\Psi_{\cM_1
\ldots \cM_n}|^2 dq_1 \ldots dq_n},
\]
and this leads to (\ref{sequential-product}).

We can also calculate $\langle p_1p_2 \ldots p_n \rangle$, the product
of the momenta of the pointers. To do this, it is convenient to move
to the momentum basis, replacing $\phi(q)$ by its Fourier transform
$\tilde \phi(p)$
and carrying out an expansion in the $p_i$:
\bea \label{p-expansion}
\Psi_{\cM_1 \ldots \cM_n}&=&\bra{\psi_f}\left(U_{n+1}e^{-igp_nA_n}U_n \ldots U_2e^{-igp_1A_1}U_1\right)\ket{\psi_i} \tilde \phi(p_1) \ldots
\tilde \phi(p_n),  \\\nonumber
&=&\bra{\psi_f}U_{n+1}U_n \ldots U_1\ket{\psi_i} \ \left( 1 -ig\sum_i
p_i(A_i)_w + (-ig)^2\sum_{i<j} p_ip_j(A_j,A_i)_w + \ldots
\right)\tilde \phi(p_1) \ldots \tilde \phi(p_n).\nonumber
\eea
Assumption A implies that only the terms in $p_1p_2 \ldots p_n$ in
$|\Psi_{\cM_1 \ldots \cM_n}|^2$ need be considered in calculating
\begin{align}
\langle p_1p_2 \ldots p_n \rangle= \frac{\int
\overline{\Psi_{\cM_1 \ldots \cM_n}} p_1 \ldots p_n \Psi_{\cM_1
\ldots \cM_n} dp_1 \ldots dp_n}{\int |\Psi_{\cM_1 \ldots \cM_n}|^2
dp_1 \ldots dp_n}.
\end{align}
It is simplest to treat the cases of $n$ even and odd separately. For
the even case we have
\be \label{even-p-product} \langle p_1p_2 \ldots p_{2m}
\rangle=2(-1)^m (gv)^{2m} \ Re \sum_{r \ge s} \sum_{\bf
i,j} (-1)^r(A_{i_r}, \ldots ,A_{i_1})_w\overline{(A_{j_s}, \ldots
, A_{j_1})}_w, \ee
and for the odd case:
\begin{align} \label{odd-p-product}
\langle p_1p_2 \ldots p_{2m+1}
\rangle=2(-1)^{m+1} (gv)^{2m+1} \ Im \sum_{r > s} \sum_{\bf
i,j} (-1)^r(A_{i_r}, \ldots ,A_{i_1})_w\overline{(A_{j_s}, \ldots ,A_{j_1})}_w,
\end{align}
where $v=\int p^2{\tilde \phi}^2(p)dp$.

The case of mixed products of positions and momenta are treated
similarly, and they depend only on the real or imaginary parts of the
sequential weak values given by (\ref{sequential-value}). For example,
to calculate $\langle q_1p_2 \rangle$ we express the first variable in
the position basis and the second in the momentum basis:
\[
\Psi_{\cM_1,\cM_2}=\bra{\psi_f}U_3U_2U_1\ket{\psi_i} \
\left( \phi(q_1)\tilde \phi(p_2)+g(A_1)_w\phi^\prime(q_1)\tilde \phi(p_2)-ig(A_2)_w\phi(q_1)p_2\tilde \phi(p_2)
+ig^2(A_2,A_1)_w\phi^\prime(q_1)p_2\tilde \phi(p_2) \right),
\]
which yields (\ref{mixed}). For these mixed products, since there
is a factor of $i$ for each $p$ in the product, we take the
imaginary part of weak values when there is an odd number of $p$'s
present and the real part otherwise.

Thus all possible expectations of products of position or momentum can
be obtained from the sequential weak values.

\end{document}